# Chain-of-Programming: Empowering Large Language Models for Geospatial Code Generation

Shuyang Hou[a], Haoyue Jiao[b], Zhangxiao Shen[a], Jianyuan Liang[a], Anqi Zhao[a], Xiaopu Zhang[a], Jianxun Wang[a], Huayi Wu[a]*

*a. State Key Laboratory of Information Engineering in Surveying, Mapping, and Remote Sensing, Wuhan University, Wuhan, China

b. School of Resource and Environmental Sciences, Wuhan University, Wuhan, China

*Corresponding author: Huayi Wu, email: <u>wuhuayi@whu.edu.cn</u>

**Abstract**

With the rapid growth of interdisciplinary demands for geospatial modeling and the rise of large language models (LLMs), geospatial code generation technology has seen significant advancements. However, existing LLMs often face challenges in the geospatial code generation process due to incomplete or unclear user requirements and insufficient knowledge of specific platform syntax rules, leading to the generation of non-executable code, a phenomenon known as "code hallucination." To address this issue, this paper proposes a Chain of Programming (CoP) framework, which decomposes the code generation process into five steps: requirement analysis, algorithm design, code implementation, code debugging, and code annotation. The framework incorporates a shared information pool, knowledge base retrieval, and user feedback mechanisms, forming an end-to-end code generation flow from requirements to code, without the need for model fine-tuning. Based on a geospatial problem classification framework and evaluation benchmarks, the CoP strategy significantly improves the logical clarity, syntactical correctness, and executability of the generated code, with improvements ranging from 3.0% to 48.8%. Comparative and ablation experiments further validate the superiority of the CoP strategy over other optimization approaches and confirm the rationality and necessity of its key components. Through case studies on building data visualization and fire data analysis, this paper demonstrates the application and effectiveness of CoP in various geospatial scenarios. The CoP framework offers a systematic, step-by-step approach to LLM-based geospatial code generation tasks, significantly enhancing code generation performance in geospatial tasks and providing valuable insights for code generation in other vertical domains.

1. Introduction

The rapid advancements in remote sensing and sensor network technologies have significantly enhanced geospatial data acquisition and processing, resulting in a dramatic increase in data volume(Breunig et al., 2020; Zhang et al., 2022a). Concurrently, the growing importance of spatiotemporal data in fields such as ecology, transportation, politics, and military affairs reflects broader trends in global defense modernization, space exploration, and competition in Earth observation(Bennett, 2020). Researchers are leveraging geospatial modeling to manage vast datasets, supporting high-precision analysis, knowledge discovery, and sustainable development while addressing the increasing demand for cross-disciplinary data analysis(Shen et al., 2023; Zhang and Wang, 2024). However, the heterogeneity, complex formats, and large scale of geospatial data present substantial challenges for the rendering and visualization capabilities of traditional compilation platforms(Soille et al., 2018; Wu et al., 2024). In response, cloud-based geospatial analysis platforms (e.g., GEE, ArcGIS, and PIE Engine) are being increasingly deployed for complex modeling tasks, offering powerful programming engines, while local tools (e.g., Python's GDAL, ArcPy, and R's Raster and Terra) facilitate fundamental geospatial modeling(Alam et al., 2022; Kempeneers et al., 2019; Liang et al., 2023; Palomino et al., 2017). The integration of cloud platforms with local tools has given rise to the emerging field of "geospatial code(Bebortta et al., 2020; Hou et al., 2024c)." Driven by the need for massive data processing and complex modeling, automated geospatial code generation research has emerged . Its primary objective is to translate natural language requirements into executable code, standardize modeling workflows, lower technical barriers for non-expert users, enhance research efficiency, and foster interdisciplinary applications(Mansourian and Oucheikh, 2024).

The automated generation of geospatial code represents a critical approach to enhancing the efficiency of geospatial modeling(Hou et al., 2024a). However, early technological paradigms struggled to achieve this goal directly, relying instead on geographical information services (GIServices) to indirectly optimize modeling processes(Li et al., 2020). These methods were based on predefined, standardized geospatial data analysis modules that required manual assembly or configuration by experts(Zipf and Jöst, 2006). The interoperability of these modules was facilitated by standards such as those defined by the Open Geospatial Consortium (OGC), enabling seamless integration and automated spatiotemporal data processing(Castronova et al., 2013; Jeppesen et al., 2018). Nevertheless, the construction of GIServices was often constrained to specific tasks or platforms, dependent on symbolic rules or data annotations, with limited generalizability(Kokla and Guilbert, 2020). Moreover, their usage heavily relied on expert judgment, making the modeling process complex and labor-intensive. As a result, geospatial modeling remained far from fully automated(El Jaouhari et al., 2024). The advent of attention mechanisms and Transformer models has led to the development of large language models (LLMs) trained on extensive corpora, endowing them with capabilities such as instruction following, contextual learning, logical reasoning, and content generation(Minaee et al., 2024; Zhao et al., 2023). These advancements have made it possible to directly translate natural language into executable code(Liu et al., 2024b). Finetuned general-purpose code generation models, such as Code LLaMA(Roziere et al., 2023), DeepSeek Coder(Guo et al., 2024), and WizardCoder(Luo et al., 2023), have demonstrated excellent performance in generic code generation tasks. However, evaluation studies indicate that these models fall short in the specialized domain of geospatial code generation, frequently exhibiting "code hallucination" phenomena that result in non-functional code(Gramacki et al., 2024; Hou et al., 2024b), as illustrated in Figure 1.

The "code hallucination" phenomenon can be attributed to several factors. First, unclear or incomplete requirement descriptions often hinder accurate code generation(Le et al., 2020). In geospatial tasks, users frequently omit critical details such as time, spatial, or dataset dimensions, or fail to specify the programming language and platform clearly, leading to misunderstandings by LLMs and the generation of incorrect code. Second, the lack of specialized syntax knowledge contributes significantly to the issue(Hou et al., 2024a). Geospatial code is often encapsulated within higher-level abstractions by cloud platforms or open-source libraries, making its control flow, data flow, and platform-specific syntax distinct from general-purpose code. These variations necessitate specialized training to apply the correct syntax and function names. Additionally, remote sensing imagery in geospatial analysis platforms involves proprietary file paths, and inadequate knowledge of these paths can result in errors during image selection and path encoding. Lastly, while prior research has sought to enhance geospatial code generation through fine-tuning, this process is resource-intensive and costly(Zhang et al., 2024a). The impact of varying corpus proportions remains underexplored, and if the corpus is imbalanced, of low quality, or if training strategies are suboptimal, it can lead to model degradation or catastrophic forgetting(Zhang et al., 2023).

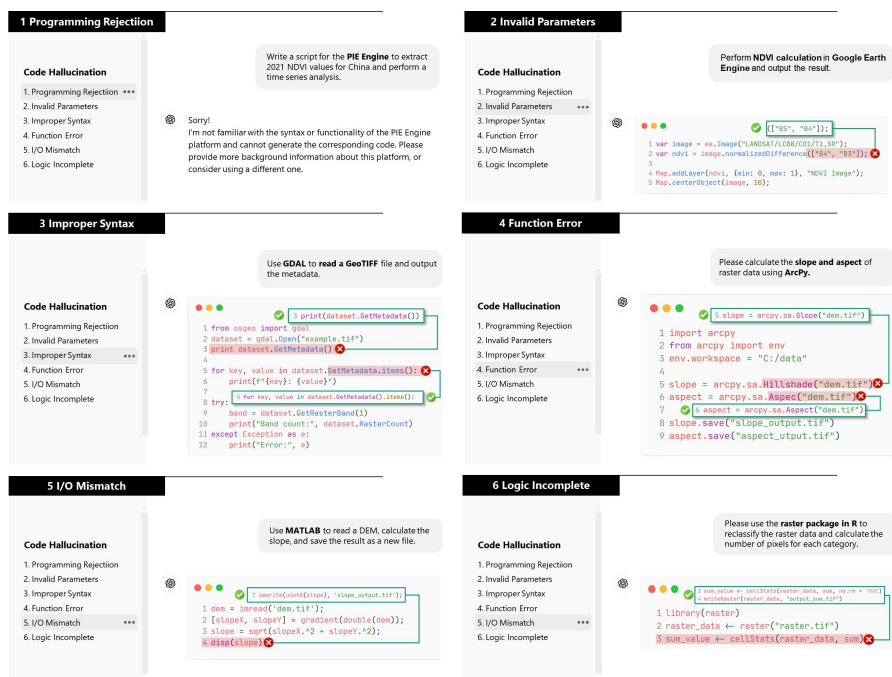

Figure 1 Illustration of the "code hallucination" phenomenon.

Although existing research and engineering models have contributed valuable insights into addressing the "code hallucination" phenomenon in geospatial code generation, they exhibit several limitations. First, the Chain-of-Thought (CoT) strategy, which excels in complex reasoning, enhances self-debugging and collaborative capabilities(Zhang et al., 2022b). However, CoT primarily activates existing knowledge, whereas geospatial code generation relies heavily on structured syntax and domain-specific knowledge—elements that current models lack. Step-by-step reasoning cannot fully mitigate this gap and may even exacerbate hallucination issues(Sprague et al., 2024). Second, the waterfall model, which divides the development process into distinct phases—requirement analysis, design, coding, testing, and maintenance—offers a standardized approach beneficial for regulated development(George, 2014). However, it fails to accommodate unclear requirements and lacks provisions for user feedback and code maintenance, which are critical for improving code accuracy and readability. Finally, Service-Oriented Architecture (SOA) and Model-as-a-Service (MaaS) treat geospatial modeling requirements as the decomposition and reorganization of service components(Mazzetti and Nativi, 2024). While this modular approach enables the construction of a complete modeling process, SOA and MaaS focus only on Requirement-to-Model mapping and lack a Model-to-Code connection, making it difficult to establish an end-to-end linkage from requirements to code.

This paper proposes a Chain of Programming (CoP) approach to address challenges such as ambiguous requirement expressions, insufficient knowledge of specific coding syntax, and the high cost and uncertain efficacy of model fine-tuning, as well as limitations in existing solutions. The CoP framework aims to improve the transformation of requirements into abstract models and bridge the gap between models and executable code, establishing an end-to-end pathway from requirements to code. Specifically, CoP structures geospatial code generation as a sequential process, divided into stages of requirement analysis, algorithm design, code implementation, testing, and maintenance. In the requirement analysis stage, CoP helps users refine their requirements, establishes contextual dependencies via a shared information pool, utilizes knowledge base retrieval to apply geospatial syntax, and incorporates expert feedback for experiential insights. Through a custom geospatial problem classification framework and a code generation evaluation benchmark, CoP enhances code readability, logical coherence, and syntactic accuracy, facilitating the generation of executable geospatial code without the need for further model fine-tuning.

In summary, the contributions of this paper are as follows:

- We propose a Chain of Programming (CoP) framework tailored for geospatial code generation with LLMs.

- We design mechanisms such as a shared information pool, knowledge base retrieval, expert feedback, and classification-based evaluation to support the effective operation of CoP.

- We establish an end-to-end bridge for Requirement-to-Code transformation, enabling seamless code generation from user requirements.

- We validate CoP's applicability in geospatial scenarios through case studies on building data visualization and fire data analysis.

- We offer a systematic approach to geospatial code generation, providing insights applicable to code generation tasks in other specialized fields.

The remainder of this paper is organized as follows: Section 2 reviews existing research on LLMs code generation, inference enhancement strategies, and related work on geospatial model building and code generation. Section 3 outlines the overall framework of the Chain of Programming (CoP) approach and data preparation. Section 4 details the evaluation framework, ablation studies, and comparative analysis of results. Section 5 presents two case studies to demonstrate practical applications. Section 6 concludes the paper and discusses limitations.

## 2. Related Work

### 2.1. Code Generation with Large LLMs

Code generation tasks focus on converting natural language inputs into source code (NL2Code)(Dehaerne et al., 2022; Jiang et al., 2024). Early approaches in this field relied on heuristic rules(Nymeyer and Katoen, 1997), expert systems(Depradine, 2003), or reinforcement learning methods(Le et al., 2022)—such as probabilistic grammar frameworks and small-scale pretrained models—that were task-specific and programming-language-dependent, with limited applicability to generating executable code(Chen et al., 2021; Jiang et al., 2024).Recent advancements in LLMs based on the Transformer architecture, trained on heterogeneous corpora (e.g., text, web content, and code), have significantly improved capabilities in contextual understanding, logical reasoning, and instruction execution, revolutionizing code generation methodologies. These models enable users to provide direct instructions, which are parsed and translated into code. However, the relatively low proportion of code in training datasets often leads to issues such as refusal to generate or producing hallucinated (nonexistent or erroneous) outputs(Agarwal et al., 2024; Liu et al., 2024a; Zhang et al., 2024b).Current optimization strategies for LLM-based code generation encompass retrieval-augmented generation, fine-tuning, and agent-based approaches(Jin et al., 2024a). Retrieval-augmented generation constructs code-specific knowledge bases to enhance model outputs by supplementing missing information without modifying the model's structure(Lewis et al., 2020). While this method improves code readability and executability, its performance heavily relies on the quality of the knowledge base and retrieval algorithms, often resulting in limited consistency and effectiveness(Gao et al., 2023).Fine-tuning on general code and task-specific datasets has led to the development of LLMs optimized for code generation (e.g., Code LLaMA, DeepSeek Coder, WizardCoder)(Chen et al., 2024). However, this approach requires large-scale datasets and computationally expensive hardware. Moreover, the lack of quantitative evaluation regarding dataset size and composition increases the risk of model performance degradation or loss of specialized knowledge. Agent-based strategies involve integrating LLMs into collaborative pipelines by assigning roles such as programmer or engineer to improve code generation performance(Huang et al., 2023). While these approaches capitalize on the strengths of LLMs, they entail significant resource consumption and communication overhead(He et al., 2024). Additionally, they often lack domain-specific expert feedback, particularly for geospatial code.In summary, current optimization strategies are predominantly designed for general-purpose code, leaving geospatial code generation constrained by issues of readability, operational reliability, and precision. These challenges highlight the critical need for targeted optimization techniques tailored to the unique requirements of geospatial applications.

**2.2. Incremental Reasoning Optimization Strategies for LLMs**

Geospatial code generation tasks, which intersect multiple disciplines such as computer programming, geographic information science, and data science, require mapping users' unstructured requirements into standardized geospatial modeling workflows and generating compliant code. This process involves complex reasoning(Breunig et al., 2020). Studies indicate that LLMs perform better on complex reasoning tasks when these tasks are broken down into smaller steps, allowing for incremental reasoning to improve accuracy(Jin et al., 2024b). Existing stepwise reasoning strategies include chain-of-thought (CoT) prompting, which decomposes reasoning steps and is suitable for logical reasoning or multi-step computation(Su et al., 2023; Zhang et al., 2022b). However, geospatial code generation workflows are highly diverse and lack fixed patterns, making it challenging to establish generalizable prompt templates(Sprague et al., 2024). Tree-of-thought (ToT) and graph-of-thought (GoT) approaches are better suited for structured tasks like knowledge graph construction, yet the complex structures in ToT often lead to information overload in programming, affecting logical coherence(Lei et al., 2023; Long, 2023). Progressive hint prompting (PHP) iteratively guides models and is useful for multi-turn task updates, but it may result in ineffective feedback when requirements are unclear. Self-improvement and self-debugging optimize output quality through feedback, but reliance on expert input reduces generation efficiency(Zheng et al., 2023). Skeleton-of-thought (SoT), minimal decomposition, and chain-of-action (CoA) approaches favor modular generation, though insufficient information flow between modules can create "information silos(Ning et al., 2023; Pan et al., 2024)." Self-collaboration methods simulate teamwork to enhance output but rely on communication mechanisms, resulting in higher resource consumption and reasoning delays(Dong et al., 2024). Currently, no reasoning optimization strategy is specifically designed to enhance LLM performance in geospatial code generation.

## 2.3. Geospatial Model Construction and Code Generation

Geospatial services are modular information services delivered over networks to access, process, analyze, and display geospatial data, facilitating data sharing and reuse(Granell et al., 2010). Geospatial modeling uses mathematical and computational techniques to simulate and predict spatial data, aiding in the understanding and forecasting of geographic phenomena(Yang et al., 2011). As the core of geospatial services, geospatial modeling supports the entire process from data acquisition to analysis, prediction, and decision-making(Breunig et al., 2020). Early geospatial modeling relied on OGC Web Services, SOAP(Suda, 2003), and REST protocols(Neumann et al., 2018), assembling workflows through modular components to meet specific analytical needs(Castronova et al., 2013). This approach typically required manual coding and intervention, limiting efficiency and flexibility, and struggled to meet diverse application demands. With the advancement of cloud technology, platforms such as Google Earth Engine (GEE), ArcGIS, and PIE Engine have integrated vast geospatial datasets with powerful computational resources(Elmahal and Ganwa, 2024). These platforms, through language-based open APIs, allow users to simplify data access and model development. Additionally, the geospatial community has developed toolkits for mainstream programming languages (e.g., GDAL and ArcPy for Python, Raster and Terra for R), enhancing the ease of basic geospatial modeling(Hou et al., 2024b). However, cloud platforms and open-source toolkits are often secondary wrappers based on foundational languages, with control and data flows differing from general code, and platform-specific function naming and parameter syntax. Proprietary formats for built-in remote sensing imagery paths further raise the entry barrier for non-specialists. The introduction of LLMs represents a transformative shift for geospatial modeling from modular assembly to automated code generation. Using natural language prompts, LLMs (e.g., GPT models) can directly generate geospatial code that aligns with user needs, achieving efficient Requirement-to-Model mapping(Hou et al., 2024a). However, the connection between model-generated code and specific implementation (Model-to-Code) remains underdeveloped, hindering fully automated Requirement-to-Code transformation. Although manual intervention is reduced, accuracy and usability challenges persist. Currently, geospatial code generation lacks tailored optimization strategies for LLMs, underscoring the need for refined prompting and knowledge-guided generation to bridge the gap from Requirement-to-Code effectively.

## 3. Chain of Programming

## 3.1. Overall Framework

The overall framework of CoP consists of five interconnected chain-like processes: requirement analysis, algorithm design, code implementation, code debugging, and code annotation, resembling the structured operational flow of the waterfall model. Each stage facilitates the transfer and sharing of structured information during the code development process through a shared information pool. To address knowledge gaps during model operation, the framework is equipped with retrieval-based knowledge bases, including a platform or toolkit knowledge base, a function syntax knowledge base, and a built-in dataset knowledge base. The complete chain-like framework is illustrated in Figure 2.

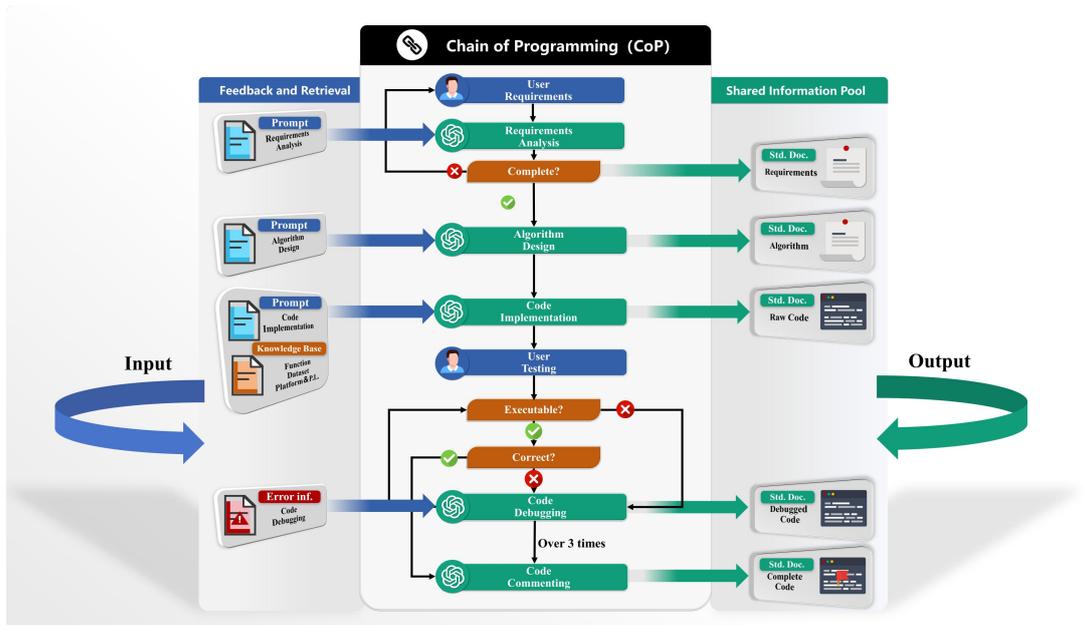

Figure 2 Schematic diagram of the Chain of Programming (CoP) framework.

First, a shared information pool is initialized as an explicit short-term memory module, serving as a central communication platform to store standardized information across different stages. This design ensures precise and efficient access to necessary information at each stage, minimizing reliance on stochastic memory, reducing redundant data retrieval, and preventing the omission of critical details. The information pool is cleared after each code generation task to provide sufficient storage capacity for subsequent tasks.

During the requirement analysis stage, users input their geospatial code programming requirements. The LLMs evaluates the completeness of the requirements using a predefined prompt template, ensuring inclusion of eight key elements: analysis platform, programming language, analysis objective, spatial extent, temporal extent, data source and format, analytical methodology, and output format and type. Among these, the analysis platform, programming language, analysis objective, data source and format, and output format and type are categorized as required items; spatial and temporal extents are conditional items, depending on the task; and the analytical methodology is considered an optional item.The model first ensures the completeness of the required items. For conditional items, the model determines their necessity based on the task type, mandating them if required or omitting them otherwise. For optional items, the model prioritizes user-provided information and supplements missing details through inference. In cases of incomplete information, the model prompts the user for clarification while filling in missing optional details as needed. Once the requirements are verified as complete, a standardized JSON-format user requirement document is generated and stored in the shared information pool. The requirement analysis prompt template and standardized information are illustrated in Figure 3.

Figure 3 Prompt engineering template for requirement analysis.

During the algorithm design stage, the standardized user requirement document is retrieved from the shared information pool and parsed to construct a multi-step algorithmic workflow. Following the divide-and-conquer principle, complex requirements are decomposed into logically independent and functionally specific submodules. This ensures that each module performs a single function, maintains a clear input-output relationship, and offers high composability. Each module includes a functional description, input specifications, output definitions, and implementation details. The finalized workflow is stored in JSON format within the shared information pool and simultaneously displayed on the console, providing a clear pathway for subsequent code generation. The prompt template and standardized information for algorithm design are illustrated in Figure 4.

## Prompt Engineering for Algorithm Design

### Task Description

- Currently in the **Algorithm Design** phase.
- Your task is to retrieve the *[ User Requirements Document ]* from the shared information pool and translate it into a multi-step algorithmic workflow.
- Following the divide-and-conquer principle, decompose complex requirements into well-defined modules, each with a **single function**, clear **input-output relations**, and **high composability**.
- Produce a structured **JSON-format** *[ Algorithm Design Document ]* and store it in the **shared information pool**.

### Step by Step

**1 Extract and analyze requirements**

Extract the standardized user requirements from the shared information pool and parse them to construct a multi-step algorithmic workflow.

**2 Decompose into independent modules**

Following the divide-and-conquer principle, decompose complex requirements into logically independent functional modules. Ensure each module includes:
- **Module_Description:** Core responsibilities and purpose of the module
- **Input:** Module input parameters
- **Output:** Expected module outputs
- **Implementation_Details:** Key logic and steps for implementation

**3 Generate standardized JSON output document.**

Once the algorithm design is completed, generate a standardized JSON requirements document.

### Expected Output Format

**1 Standardized Algorithm Design Document**

```json
{
  "Document_type": "Algorithm Design Document",
  "Algorithm": [
    {
      "Module_Sequence": 1,
      "Module_Name": "Data Acquisition",
      "Module_Description": "Retrieve land use data for a specified city and time period",
      "Input": "City name, time range",
      "Output": "Geospatial dataset containing land use information",
      "Implementation_Details": "Access API or database to obtain GeoJSON files relevant to the city and time period"
    },
    {
      "Module_Sequence": 2,
      "Module_Name": "Data Cleaning and Preprocessing",
      "Module_Description": "Clean and standardize data to ensure usability",
      "Input": "Acquired geospatial dataset",
      "Output": "Cleaned dataset with uniform format and standardized fields",
      "Implementation_Details": "Validate and convert data format (e.g., GeoJSON to Shapefile or GeoDataFrame); handle missing or anomalous values"
    },
    {
      "Module_Sequence": 3,
      "Module_Name": "Spatial Analysis (Area Calculation)",
      "Module_Description": "Calculate area for each land type",
      "Input": "Cleaned dataset",
      "Output": "Area statistics for each land type",
      "Implementation_Details": "Use GIS tools (e.g., GeoPandas, Shapely) to compute area by land type, storing results in a structured format"
    },
    {
      "Module_Sequence": 4,
      "Module_Name": "Temporal Comparison Analysis",
      "Module_Description": "Compare land use data over different time periods to calculate changes in area by land type",
      "Input": "Area data for initial and target time periods",
      "Output": "Change in area for each land type",
      "Implementation_Details": "Use time-series comparison algorithms to calculate change rates and generate tables of area changes"
    },
    {
      "Module_Sequence": 5,
      "Module_Name": "Generate Statistical Report",
      "Module_Description": "Compile analysis results into a report on land use area changes by type",
      "Input": "Area change data",
      "Output": "Statistical report with land use area changes",
      "Implementation_Details": "Use templated report generation tools or scripts to create readable report files with charts and tables (e.g., PDF or HTML)"
    }
  ]
}
```

Figure 4 Prompt engineering template for algorithm design

During the code implementation stage, the standardized user requirements and algorithmic workflow stored in the shared information pool are used to generate the code. If operator or syntax details are required, the model can query the built-in dataset knowledge base and function syntax knowledge base. The generated code is stored in the shared information pool in .txt format. The prompt template and standardized information for code implementation are illustrated in Figure 5.

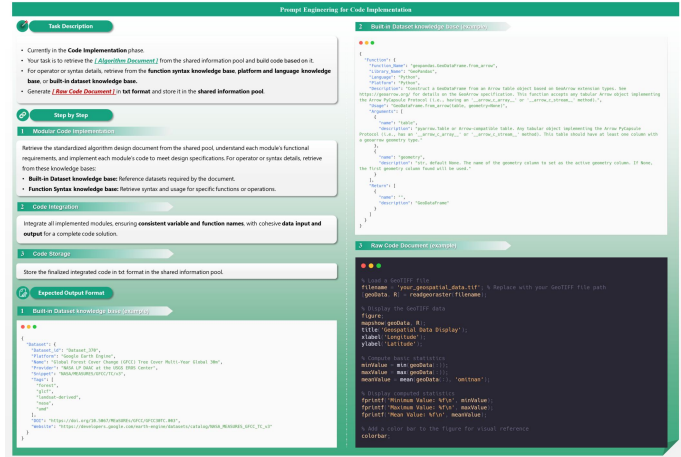

Figure 5 Prompt engineering template for code implementation.

The code debugging and commenting stages are conducted iteratively. During the debugging stage, users execute the code and provide feedback on its functionality, including executability (Y/N) and result accuracy (Y/N). The process operates as follows: if the code is executable and produces correct results, it proceeds to the commenting stage; if it is not executable, the console error message is recorded, and the debugging process restarts; if it is executable but the results are incorrect, feedback on the unexpected output is provided through console interaction, and the debugging process resumes for correction. The maximum number of debugging iterations is controlled by a hyperparameter, with a default limit of three iterations. The debugging process repeats the state evaluation until a "Y/Y" feedback is achieved or the iteration limit is reached, after which the process transitions to the commenting stage. During the commenting stage, metadata such as the creation time, applicable platform, and a brief description of the code's overall functionality is added at the beginning of the script. Comments must adhere to the syntax standards of the programming language used, with at least one concise and clear comment per module. Existing comments are optimized to ensure logical consistency throughout the code. Finally, the complete code is stored in .txt format. The prompt template and standardized information for code implementation are illustrated in Figure 6.

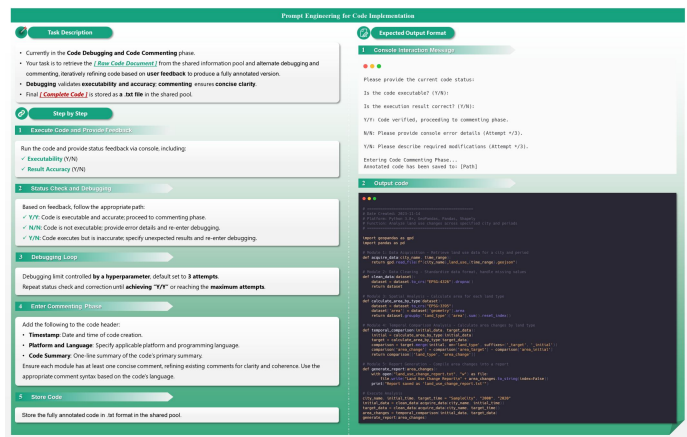

Figure 6 Prompt engineering template for code debugging and commenting.

### 3.2. Data Preparation

### 3.2.1. Retrieval Knowledge Bases

The retrieval knowledge bases include a platform or toolkit knowledge base, a function syntax knowledge base, and a built-in dataset knowledge base, all stored in JSON format. The platform or toolkit knowledge base was compiled by experts using geospatial platform-related information sourced from Wikipedia, official documentation, GitHub, and technical forums such as Stack Overflow. It contains 14 records covering platform descriptions, types, task suitability, data source interfaces, access permissions, technical support, and cross-platform compatibility.The function syntax knowledge base draws from official documentation of geospatial platforms and function libraries, including materials from platform websites and GitHub repositories. It contains 8,729 records with fields such as Operator_id, Full_name, Short_name, Library_name, Language, Platform, Description, Usage, Parameters, and Output_type.The built-in dataset knowledge base aggregates preprocessed remote sensing datasets from platforms such as GEE and PIE Engine, enabling users to directly access datasets via indexing. It includes 2,732 records with fields such as Dataset_id, Name, Provider, Snippet, Tags, Description, DOI, and Website.Details of the data collection process are summarized in Table 1.

**Table 1 Data Collection for Knowledge Bases**

| Category | Download Source | Quantity (Items) |
| --- | --- | --- |
| Platform or Toolkit | Wikipedia, Official Platform Documentation, GitHub Pages, Stack Overflow | 14 |
| Function Syntax | Official Platform Documentation, Official Toolkit Website, GitHub Pages | 8729 |
| Built-in Dataset | Official Platform Documentation (GEE and PIE Engine) | 2732 |

### 3.2.2. Evaluation Datasets

The evaluation dataset employs the GeoCode-Eval benchmark proposed in this study(Hou et al., 2024b). This benchmark is constructed from open-source question banks, documentation, and real user code files, generated through expert input and LLMs using the Self-Instruct framework. GeoCode-Eval has been utilized in three previous studies to assess the geospatial code generation capabilities of LLMs(Hou et al., 2024a; Hou et al., 2024b; Hou et al., 2024c).The dataset includes eight categories of tasks: operator knowledge, dataset knowledge, platform or toolkit knowledge, platform or toolkit identification, programming language identification, entity recognition, code summarization, and code generation, comprising a total of 4,000 questions. From this dataset, 500 code generation tasks were selected as the basis for this study and further categorized into three major groups and eight subcategories, as detailed in Table 2.

**Table.2 Data Collection for Evaluation Datasets**

| Id | Primary Category | Secondary Category | Task Description | Examples |
| --- | --- | --- | --- | --- |

| Id | Primary Category | Secondary Category | Task Description | Examples |
|---|---|---|---|---|
| 1 | **Data Preparation and Preprocessing** | Geometry and Area Definition & Data Extraction | Includes tasks such as defining geometric objects, setting spatial extents, and extracting data for specific regions | Define city boundaries, extract meteorological data for a specific area |
| 2 | | Image and Raster Data Processing | Tasks such as image clipping, reprojection, filtering, band calculation, and image stacking | Convert multi-band images to single-band, perform radiometric correction |
| 3 | | Spatiotemporal Analysis & Data Aggregation | Involves time- and space-based analysis tasks, including spatiotemporal aggregation, trend analysis, and data statistics | Calculate city temperature time series, analyze precipitation trends in a specific area |
| 4 | **Data Analysis** | Vegetation Indices & Environmental Metrics Calculation | Includes calculation of NDVI, EVI, and other vegetation and environmental metrics | Calculate NDVI or EVI for a region, compute PM2.5 concentration mean |
| 5 | | Land Cover and Classification | Encompasses land cover classification and land-use change detection tasks | Generate land-use classification maps, monitor land cover changes |
| 6 | | Hydrology and Meteorology Analysis | Covers hydrological (e.g., watershed delineation, runoff calculation) and meteorological data analysis (e.g., precipitation, temperature) | Generate watershed precipitation distribution, calculate annual average rainfall |
| 7 | | Data Export and Format Conversion | Includes exporting data in specific formats (e.g., CSV, GeoTIFF) and format conversions | Export NDVI data for a region, convert raster data to vector format |
| 8 | **Data Output and Visualization** | Visualization and Chart Generation | Covers tasks related to visualizing spatial data, such as map, chart, and time series generation | Display vegetation cover change map, generate PM2.5 concentration time series chart |

To ensure the accuracy of entity recognition, questions undergo necessary entity annotation, including user requirements, platform, programming language, analysis objectives, spatial extent, temporal extent, data source and format, analytical methodology, and output format and type. Figure 7 illustrates user requirements and corresponding standard answers for requirement analysis across different types of tasks.

Figure 7 Example of Evaluation Set Questions

## 4. Evaluation

The evaluation aims to scientifically quantify the following issues across various mainstream LLMs:

**EQ1**: The performance change of the CoP strategy compared to zero-shot performance.

**EQ2**: The performance change of the CoP strategy compared to other optimization strategies.

**EQ3**: The necessity of the three key mechanisms in the CoP strategy: shared information pool, knowledge retrieval system, and expert feedback.

**EQ4**: The performance change between multiple CoP strategy fine-tuning and multiple zero-shot fine-tuning iterations.

### 4.1. Baseline Models

The CoP benchmark models selected in this study include mainstream high-performance commercial closed-source models developed after 2023, as well as general-purpose open-source LLMs and code generation models. The CoP strategy is applied to these models to compare the accuracy of code generation in a zero-shot setting. Detailed information about the baseline models is provided in Table 3.

Table 3 Baseline Model Information Table

| Category | Model Name | Company | Size | Date | Open Source |
|---|---|---|---|---|---|
| Commercial LLMs | GPT-4 | OpenAI | N/A | 2023 | No |
| | GPT-4-turbo | OpenAI | N/A | 2024 | No |
| | Claude-3-Opus | Anthropic | N/A | 2023 | No |
| | ERNIE-4.0 | Baidu | N/A | 2023 | No |
| General LLMs | LLaMA3 | Meta AI | 70B | 2024 | Yes |
| | PaLM | Google | 540B | 2022 | No |
| | BLOOM | BigScience | 176B | 2022 | Yes |
| Code-Generation LLMs | CodeGemma | Google | 7B | 2023 | Yes |
| | StarCoder 2 | BigCode | 15B | 2023 | Yes |
| | CodeQwen | Alibaba | 14B | 2023 | Yes |
| | WizardCoder | WizardLM | 15B | 2023 | Yes |
| | Code Llama | Meta AI | 13B | 2023 | Yes |
| | OctoCoder | Hugging Face | 15.5B | 2023 | Yes |
| | CodeGeeX2 | THU/IDEA | 6B | 2023 | Yes |
| | CodeT5+ | Salesforce Research | 16B | 2023 | Yes |
| | CodeGen | Salesforce Research | 16.1B | 2023 | Yes |
| | CodeX | OpenAI | 12B | 2023 | Yes |

**4.2. Optimization Strategy**

The comparison strategies selected in this study are currently mainstream optimization inference strategies in the field of code generation. These strategies are applied to GPT-4, PaLM-540B, and Code Llama-13B for comparison to achieve the evaluation results. Detailed information on the optimization strategies is provided in Table 4.

Table 4 Optimization Strategy Information Table

| Category | Strategy | Year | First Institution | Authors |
|---|---|---|---|---|
| Prompt Engineering | Few-Shot | 2020 | Johns Hopkins University | Brown T. B. et al. |
| | CoT | 2022 | Google Research | Wei J., et al. |
| | Self-Edit | 2023 | Peking University | Zhang K., et al. |
| | Self-Debugging | 2023 | Google DeepMind | Chen X., et al. |

| Category | Strategy | Year | First Institution | Authors |
|---|---|---|---|---|
| Single-Agent | INTERVENOR | 2023 | Northeastern University | Wang H., et al. |
| | Self-Planning | 2024 | Peking University | Jiang X., et al. |
| | CodeCoT | 2024 | University of Hong Kong | Huang D., et al. |
| | ReAct | 2022 | Princeton University | Yao S., et al. |
| | Reflexion | 2023 | Northeastern University | Shinn N., et al. |
| | RAP | 2023 | UC San Diego | Hao S., et al. |
| | ToT | 2024 | Princeton University | Yao S., et al. |
| | CodeAgent | 2024 | Peking University | Zhang K., et al. |
| Multi-Agent | AgentCoder | 2023 | University of Hong Kong | Huang D., et al. |
| | ChatDev | 2023 | Tsinghua University | Qian C., et al. |
| | AgentVerse | 2023 | Tsinghua University | Chen W., et al. |
| | MetaGPT | 2023 | Deep Wisdom | Hong S., et al. |
| | Self-Collaboration | 2024 | Peking University | Dong Y., et al. |

### 4.3. Evaluation Results

#### 4.3.1. EQ1: Overall Performance?

The effectiveness of applying the CoP strategy to all baseline models was evaluated using four metrics: Matchability, Executability, Accuracy, and Readability. The baseline models generated code in a zero-shot manner based on the given prompts, while this study required the models, through prompt engineering, to output the corresponding entity information for the generated code, including Platform, Programming Language, Analysis Goal, Spatial Extent, Temporal Extent, Data Source and Format, Analysis Methodology, and Output Format. Under the CoP strategy, the generated code automatically incorporates these entity details. The four metrics were assessed by expert evaluations: Matchability was scored based on the entity matching rate; Executability was determined by the code's ability to run in a compiler, with the proportion of code that executed successfully (regardless of correctness) used as the executability score; Accuracy was defined by whether the execution results aligned with the expected outcomes; Finally, Readability was rated by five experts on a scale of 1 to 10, excluding the highest and lowest scores, and averaging the remaining three to determine the readability score. All scores were then converted to a percentage scale, retaining one decimal place. The specific scores for EQ1 are shown in Table 5.

**Table 5 Scores for Evaluation Metric EQ1**

| Category | Model Name | Cop? | Ma. | Exe. | Acc. | Re. |
|---|---|---|---|---|---|---|

| Category | Model Name | Cop? | Ma. | | Exe. | | Acc. | | Re. | |
|---|---|---|---|---|---|---|---|---|---|---|
| **Commercial LLMs** | GPT-4 | ✗ | 62.6 | 34.6 | 52.8 | 39.5 | 42.6 | 44.2 | 85.6 | 7.8 |
| | | ✓ | 97.2 | | 92.3 | | 86.8 | | 93.4 | |
| | GPT-4-turbo | ✗ | 60.8 | 31.4 | 53.2 | 41.3 | 43.4 | 46.2 | 84.2 | 7.4 |
| | | ✓ | 92.2 | | 94.5 | | 89.6 | | 91.6 | |
| | Claude-3-Opus | ✗ | 40.4 | 37.4 | 36.4 | 37.1 | 29.8 | 41.0 | 85.2 | 5.0 |
| | | ✓ | 77.8 | | 73.5 | | 70.8 | | 90.2 | |
| | ERNIE-4.0 | ✗ | 33.2 | 33.0 | 30.2 | 33.9 | 24.8 | 28.6 | 76 | 9.0 |
| | | ✓ | 66.2 | | 64.1 | | 53.4 | | 85 | |
| **General LLMs** | LLaMA3-70B | ✗ | 46.4 | 33.2 | 38.8 | 25.4 | 31.2 | 30.0 | 74 | 3.4 |
| | | ✓ | 79.6 | | 64.2 | | 61.2 | | 77.4 | |
| | PaLM-540B | ✗ | 43.4 | 34.8 | 47.2 | 29.2 | 40.2 | 23.6 | 77.6 | 4.6 |
| | | ✓ | 78.2 | | 76.4 | | 63.8 | | 82.2 | |
| | BLOOM-176B | ✗ | 31.6 | 35.6 | 25.6 | 41.7 | 20.4 | 36.0 | 68.8 | 6.4 |
| | | ✓ | 67.2 | | 67.3 | | 56.4 | | 75.2 | |
| **Code-Generation LLMs** | CodeGemma-7B | ✗ | 27.2 | 42.2 | 12.0 | 48.8 | 7.6 | 42.0 | 60.4 | 6.6 |
| | | ✓ | 69.4 | | 60.8 | | 49.6 | | 67 | |
| | StarCoder2-15B | ✗ | 50.6 | 30.4 | 34.0 | 42.4 | 25.8 | 27.4 | 71.6 | 3.0 |
| | | ✓ | 81.0 | | 76.4 | | 53.2 | | 74.6 | |
| | CodeQwen-14B | ✗ | 38.0 | 34.6 | 24.8 | 38.3 | 18.6 | 32.8 | 70.4 | 5.6 |
| | | ✓ | 72.6 | | 63.1 | | 51.4 | | 76 | |
| | WizardCoder-15B | ✗ | 46.8 | 23.6 | 30.2 | 34.7 | 22.6 | 30.8 | 72.4 | 7.2 |
| | | ✓ | 70.4 | | 64.9 | | 53.4 | | 79.6 | |
| | CodeLlama-13B | ✗ | 37.6 | 28.2 | 23.2 | 39.6 | 17.2 | 35.0 | 64 | 8.8 |
| | | ✓ | 65.8 | | 62.8 | | 52.2 | | 72.8 | |
| | OctoCoder-15.5B | ✗ | 43.2 | 26.2 | 26.0 | 36.2 | 19.0 | 32.0 | 58.8 | 7.8 |
| | | ✓ | 69.4 | | 62.2 | | 51.0 | | 66.6 | |
| | CodeGeeX2-6B | ✗ | 35.2 | 27.4 | 20.8 | 33.8 | 15.2 | 29.2 | 57.2 | 9.8 |
| | | ✓ | 62.6 | | 54.6 | | 44.4 | | 67 | |
| | CodeT5+-16B | ✗ | 47.4 | 35.0 | 31.6 | 31.9 | 24.0 | 26.2 | 53.2 | 7.4 |
| | | ✓ | 82.4 | | 63.5 | | 50.2 | | 60.6 | |
| | CodeGen-16.1B | ✗ | 47.6 | 36.8 | 32.4 | 35.1 | 24.6 | 29.2 | 48.8 | 5.2 |
| | | ✓ | 84.4 | | 67.5 | | 53.8 | | 54 | |
| | CodeX-12B | ✗ | 34.2 | 34.4 | 19.6 | 34.3 | 14.0 | 28.8 | 39.2 | 8.8 |
| | | ✓ | 68.6 | | 53.9 | | 42.8 | | 48 | |

The overall performance evaluation results show that the application of the CoP strategy leads to improvements in the matchability, executability, accuracy, and readability of generated geospatial code across all model types. Notably, commercial large models such as GPT-4 and GPT-4-turbo achieved scores ranging from 89.6 to 97.2, nearing their performance limits. Further analysis indicates that the CoP strategy enhances code generation performance across models of various scales. For larger models, such as GPT-4, matchability and accuracy improved from 62.6 and 52.8 to 97.2 and 86.8, respectively. For smaller models, such as CodeGemma-7B, matchability and accuracy increased from 27.2 and 12.0 to 69.4 and 49.6, while CodeGeeX2-6B saw an improvement in executability from 27.4 to 54.6. These data demonstrate that the CoP strategy exhibits strong adaptability across model scales, showcasing both universality and scalability. Additionally, the results show that the CoP strategy consistently optimizes both general-purpose LLMs and code-generation-focused LLMs. For example, in the general-purpose model LLaMA3-70B, CoP improved matchability from 46.4 to 79.6 and accuracy from 38.8 to 61.2. In the code generation-specific model StarCoder 2-15B, CoP increased executability from 30.4 to 76.4 and accuracy from 34.0 to 53.2. These findings indicate that the CoP strategy significantly enhances the structure and executability of code generation in geospatial information processing tasks and is applicable to different types of models. The visualized evaluation data is shown in Figure 8.

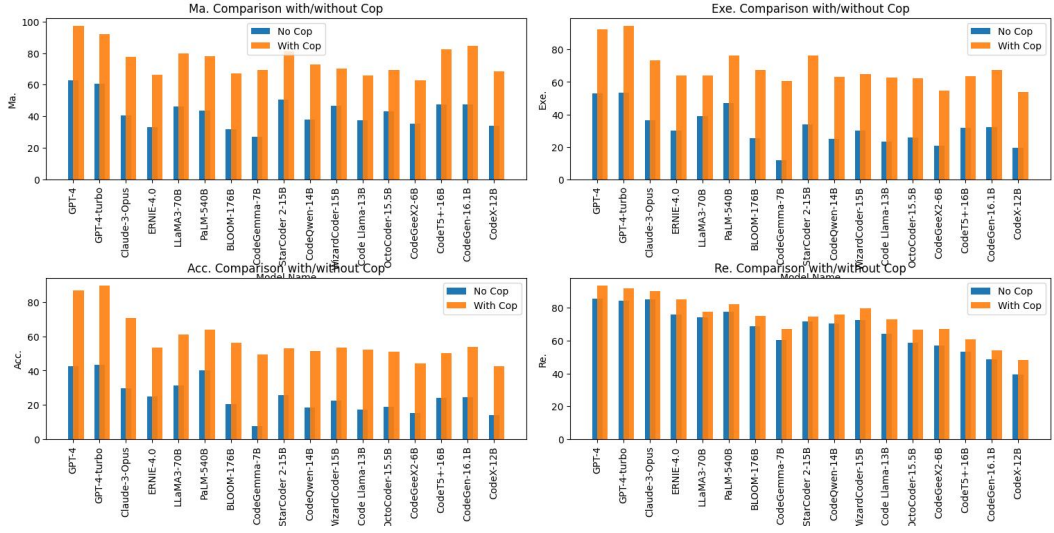

Figure 8 Improvement in Model Metrics Before and After Applying the CoP Strategy

### 4.3.2. EQ2：Compared to other optimization strategies?

This study selected three representative models from three categories: a commercial LLM (GPT-4), a general-purpose LLM (PaLM-540B), and a model focused on code generation (Code Llama-13B), to quantitatively evaluate their code generation performance under various optimization strategies. The performance metrics used for evaluation include Matchability (Ma.), Executability (Exe.), Accuracy (Acc.), and Readability (Re.), with the specific evaluation methods described in Section 4.3. The specific scores for EQ2 are shown in Table 6.

**Table 6 Scores for Evaluation Metric EQ2**

| Category | Strategy | Ma. | | Exe. | | Acc. | | Re. | |
|---|---|---|---|---|---|---|---|---|---|
| | | \multicolumn{8}{c|}{GPT-4} |
| | zero-shot | 62.6 | 0 | 52.8 | 0 | 42.6 | 0 | 85.6 | 0 |
| | Few-Shot | 60.6 | -2 | 55.4 | 2.6 | 40.2 | ↓2.4 | 82 | ↓3.6 |
| | CoT | 59.4 | ↓3.2 | 53.2 | 0.4 | 39 | ↓3.6 | 78.4 | ↓7.2 |
| **Prompt Engineering** | Self-Edit | 64.2 | 1.6 | 60.2 | 7.4 | 45.2 | 2.6 | 83 | ↓2.6 |
| | Self-Debugging | 66.0 | 3.4 | 64.2 | 11.4 | 47 | 4.4 | 81.4 | ↓4.2 |
| | INTERVENOR | 67.5 | 4.9 | 61.8 | 9 | 43.6 | 1 | 80.2 | ↓5.4 |
| | Self-Planning | 61.8 | ↓0.8 | 50.4 | -2.4 | 41 | ↓1.6 | 79 | ↓6.6 |
| **Single-Agent** | CodeCoT | 69.2 | 6.6 | 58.6 | 5.8 | 48.6 | 6 | 86.4 | 0.8 |

| Category | Strategy | Ma. | | Exe. | | Acc. | | Re. | |
|---|---|---|---|---|---|---|---|---|---|
| | ReAct | 58.4 | ↓4.2 | 49 | -3.8 | 38.4 | ↓4.2 | 77 | ↓8.6 |
| | Reflexion | 72.0 | 9.4 | 65.2 | 12.4 | 51 | 8.4 | 88 | 2.4 |
| | RAP | 73.2 | 10.6 | 67.4 | 14.6 | 52.4 | 9.8 | 87.4 | 1.8 |
| | ToT | 60.2 | ↓2.4 | 51.2 | ↓1.6 | 40.8 | ↓1.8 | 78 | ↓7.6 |
| | CodeAgent | 75.4 | 12.8 | 70 | 17.2 | 55.6 | 13 | 89.2 | 3.6 |
| | AgentCoder | 77.8 | 15.2 | 72.2 | 19.4 | 58.4 | 15.8 | 90 | 4.4 |
| | ChatDev | 79.0 | 16.4 | 75.6 | 22.8 | 60 | 17.4 | 88.6 | 3 |
| **Multi-Agent** | AgentVerse | 61.2 | -1.4 | 54 | 1.2 | 41.2 | ↓1.4 | 80.4 | ↓5.2 |
| | MetaGPT | 63.6 | 1 | 56.8 | 4 | 44.8 | 2.2 | 82.8 | ↓2.8 |
| | Self-Collaboration | 85.4 | 22.8 | 80.8 | 28 | 65.4 | 22.8 | 91 | 5.4 |
| | **CoP** | **97.2** | **34.6** | **92.4** | **39.6** | **86.8** | **44.2** | **93.4** | **7.8** |
| | | | | **PaLM-540B** | | | | | |
| | zero-shot | 43.4 | 0 | 47.2 | 0 | 40.2 | 0 | 77.6 | 0 |
| | Few-Shot | 46 | 2.6 | 49 | 1.8 | 38 | ↓2.2 | 75.8 | ↓1.8 |
| | CoT | 45.2 | 1.8 | 46.5 | ↓0.7 | 36.5 | ↓3.7 | 73.5 | ↓4.1 |
| **Prompt Engineering** | Self-Edit | 48.6 | 5.2 | 52.3 | 5.1 | 42.8 | 2.6 | 79 | 1.4 |
| | Self-Debugging | 50.4 | 7 | 53.8 | 6.6 | 44.3 | 4.1 | 78.3 | 0.7 |
| | INTERVENOR | 49 | 5.6 | 50.2 | 3 | 41 | 0.8 | 76.5 | ↓1.1 |
| | Self-Planning | 44.6 | 1.2 | 48 | 0.8 | 39 | ↓1.2 | 74.2 | ↓3.4 |
| | CodeCoT | 52.8 | 9.4 | 54.6 | 7.4 | 46.7 | 6.5 | 80.1 | 2.5 |
| | ReAct | 42.6 | ↓0.8 | 45 | ↓2.2 | 35.5 | ↓4.7 | 72 | ↓5.6 |
| **Single-Agent** | Reflexion | 54.6 | 11.2 | 56.5 | 9.3 | 48.5 | 8.3 | 81 | 3.4 |
| | RAP | 56 | 12.6 | 57.8 | 10.6 | 50.2 | 10 | 81.5 | 3.9 |
| | ToT | 44 | 0.6 | 46.8 | ↓0.4 | 38.3 | ↓1.9 | 74 | ↓3.6 |
| | CodeAgent | 57.6 | 14.2 | 59.2 | 12 | 53.5 | 13.3 | 83 | 5.4 |
| | AgentCoder | 58.8 | 15.4 | 61 | 13.8 | 55 | 14.8 | 79.2 | 1.6 |
| | ChatDev | 60 | 16.6 | 62.3 | 15.1 | 57.2 | 17 | 72.8 | ↓4.8 |
| **Multi-Agent** | AgentVerse | 45.8 | 2.4 | 48.7 | 1.5 | 39.5 | ↓0.7 | 75.2 | ↓2.4 |
| | MetaGPT | 47.2 | 3.8 | 50.5 | 3.3 | 42.3 | 2.1 | 76 | ↓1.6 |
| | Self-Collaboration | 63.6 | 20.2 | 65 | 17.8 | 60 | 19.8 | 78.5 | 0.9 |
| | **CoP** | **78.2** | **34.8** | **76.4** | **29.2** | **63.8** | **23.6** | **82.2** | **4.6** |
| | | | | **Code Llama-13B** | | | | | |
| | zero-shot | 37.6 | 0 | 23.2 | 0 | 17.2 | 0 | 64.0 | 0.0 |
| | Few-Shot | 39.2 | 1.6 | 25.6 | 2.4 | 19 | 1.8 | 64.5 | 0.5 |
| | CoT | 38 | 0.4 | 24 | 0.8 | 18.2 | 1 | 64.2 | 0.2 |
| **Prompt Engineering** | Self-Edit | 41 | 3.4 | 27.5 | 4.3 | 20.8 | 3.6 | 65.0 | 1.0 |
| | Self-Debugging | 42.5 | 4.9 | 28.8 | 5.6 | 21.5 | 4.3 | 65.2 | 1.2 |
| | INTERVENOR | 40.8 | 3.2 | 26.2 | 3 | 20 | 2.8 | 64.7 | 0.7 |
| | Self-Planning | 39 | 1.4 | 24.8 | 1.6 | 18.5 | 1.3 | 64.4 | 0.4 |
| | CodeCoT | 43.3 | 5.7 | 29.5 | 6.3 | 22.3 | 5.1 | 65.4 | 1.4 |
| | ReAct | 36.4 | ↓1.2 | 22 | ↓1.2 | 16 | ↓1.2 | 63.7 | ↓0.3 |
| **Single-Agent** | Reflexion | 44.5 | 6.9 | 30.4 | 7.2 | 23.6 | 6.4 | 65.6 | 1.6 |
| | RAP | 45.2 | 7.6 | 31.6 | 8.4 | 24.8 | 7.6 | 65.9 | 1.9 |

| Category | Strategy | Ma. | | Exe. | | Acc. | | Re. | |
|---|---|---|---|---|---|---|---|---|---|
| | ToT | 37.5 | -0.1 | 23.5 | 0.3 | 17.8 | 0.6 | 64.1 | 0.1 |
| | CodeAgent | 42 | 4.4 | 32 | 8.8 | 25.2 | 8 | 66.0 | 2.0 |
| | AgentCoder | 43.8 | 6.2 | 34.2 | 11 | 26.5 | 9.3 | 66.4 | 2.4 |
| | ChatDev | 44 | 6.4 | 35.6 | 12.4 | 27.4 | 10.2 | 66.8 | 2.8 |
| **Multi-Agent** | AgentVerse | 39.5 | 1.9 | 24.4 | 1.2 | 18.2 | 1 | 64.3 | 0.3 |
| | MetaGPT | 40.2 | 2.6 | 26.8 | 3.6 | 19.6 | 2.4 | 64.8 | 0.8 |
| | Self-Collaboration | 47 | 9.4 | 36.4 | 13.2 | 28 | 10.8 | 66.9 | 2.9 |
| | **CoP** | **65.8** | **28.2** | **62.8** | **39.6** | **52.2** | **35** | **72.8** | **8.8** |

The evaluation results indicate that the effectiveness of various optimization strategies in improving model performance varies significantly. Some strategies, such as ReAct and ToT, did not meet expectations and even had negative impacts on certain models. For example, with GPT-4, the matchability decreased from 62.6 to 58.4 and readability dropped from 85.6 to 77 under the ReAct strategy, while the matchability decreased to 60.2 under the ToT strategy. This may be due to the complexity of the reasoning process, which increased generation uncertainty and did not align well with the continuous requirements of geospatial code generation. In contrast, Agent-based strategies showed more effectiveness, especially in large-scale models. For instance, under the AgentCoder strategy, GPT-4's matchability improved from 62.6 to 77.8 and executability increased from 52.8 to 72.2, likely due to the enhanced collaborative capabilities of multiple agents in large models that better leverage knowledge representation and computational power.Meanwhile, prompt-based strategies demonstrated more notable improvements in smaller models. For example, with Code Llama-13B, the Self-Edit strategy boosted matchability from 37.6 to 41.0, suggesting that the simplicity of prompt strategies is better suited for guiding generation in smaller models. Additionally, the Self-Collaboration and ChatDev strategies showed some robustness across models of different scales, though their overall performance gains were not as significant as those achieved with the CoP strategy. The CoP strategy, which integrates prompt engineering, external knowledge base access, and expert feedback, enhanced contextual consistency and information integration. For example, with GPT-4, the CoP strategy improved matchability from 62.6 to 97.2 and executability from 52.8 to 92.4.Notably, other strategies generally led to a decrease in readability. For instance, with GPT-4, readability under the ToT strategy decreased from 85.6 to 78.4, and under the ReAct strategy, it dropped to 77. However, due to CoP's unique code-commenting mechanism, readability improved by 7.8 points. The specific numerical results of this improvement are visualized in Figure 9.

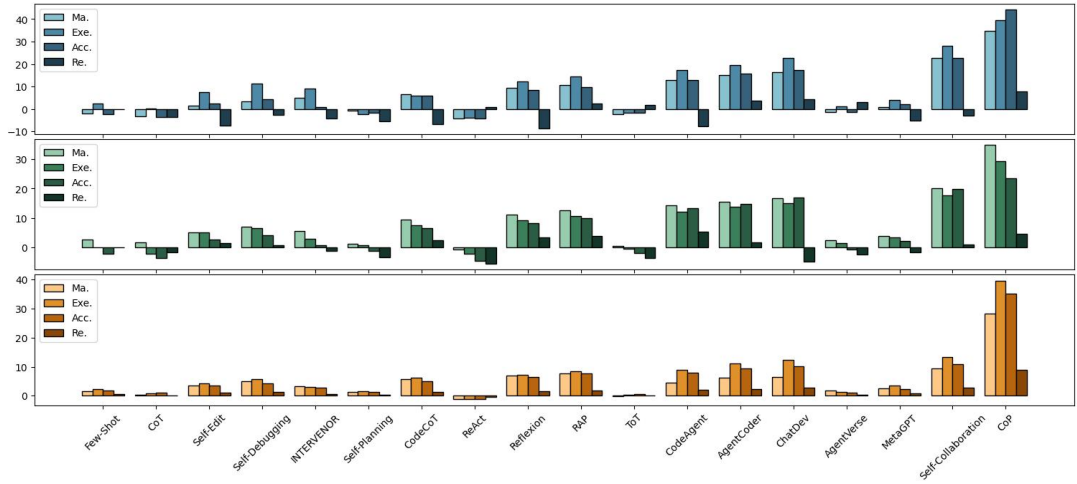

Figure 9: Improvement in Metrics for Each Optimization Strategy Compared to the Zero-Shot Strategy

**4.3.3. EQ3: Necessity of the Three Mechanisms?**

In this study, models were selected from three representative types: commercial LLMs (e.g., GPT-4), general-purpose LLMs (e.g., PaLM-540B), and models focused on code generation (e.g., Code Llama-13B). The code generation performance of these models was quantitatively evaluated under different prompting strategies. The evaluation metrics included Matchability (Ma.), Executability (Exe.), Accuracy (Acc.), and Readability (Re.), with the specific methods outlined in Section 4.3. The specific scores for EQ3 are shown in Table 7.

**Table7 Scores for Evaluation Metric EQ3**

| CoP | | | Ma. | | Exe. | | Acc. | | Re. | |
|---|---|---|---|---|---|---|---|---|---|---|
| Pool | Retrieval | Feedback | | | | | | | | |
| GPT-4 | | | | | | | | | | |
| ✗ | ✗ | ✗ | 69.3 | 27.9 | 59.7 | 32.7 | 54.2 | 32.6 | 63.8 | 29.6 |
| ✓ | ✗ | ✗ | 84.1 | 13.1 | 71.3 | 21.1 | 64.8 | 22 | 74.6 | 18.8 |
| ✗ | ✓ | ✗ | 76.4 | 20.8 | 74.8 | 17.6 | 67.5 | 19.3 | 69.9 | 23.5 |
| ✗ | ✗ | ✓ | 71.9 | 25.3 | 78.6 | 13.8 | 70.1 | 16.7 | 81.2 | 12.2 |
| ✓ | ✓ | ✗ | 89.5 | 7.7 | 85.2 | 7.2 | 75.7 | 11.1 | 84.3 | 9.1 |
| ✗ | ✓ | ✓ | 78.8 | 18.4 | 88.9 | 3.5 | 80.3 | 6.5 | 87.6 | 5.8 |
| ✓ | ✗ | ✓ | 88.3 | 8.9 | 85.7 | 6.7 | 82.5 | 4.3 | 91.4 | 2 |
| ✓ | ✓ | ✓ | **97.2** | | **92.4** | | **86.8** | | **93.4** | |
| PaLM-540B | | | | | | | | | | |
| ✗ | ✗ | ✗ | 54.3 | 23.9 | 50.7 | 25.7 | 42.5 | 21.3 | 57.4 | 24.8 |
| ✓ | ✗ | ✗ | 69.8 | 8.4 | 58.3 | 18.1 | 52.1 | 11.7 | 65.7 | 16.5 |
| ✗ | ✓ | ✗ | 60.5 | 17.7 | 64.2 | 12.2 | 48.9 | 14.9 | 61.3 | 20.9 |
| ✗ | ✗ | ✓ | 57.6 | 20.6 | 66.1 | 10.3 | 53.7 | 10.1 | 67.2 | 15 |
| ✓ | ✓ | ✗ | 72.1 | 6.1 | 71.8 | 4.6 | 57.3 | 6.5 | 74.5 | 7.7 |

|  | CoP |  | Ma. |  | Exe. |  | Acc. |  | Re. |  |
| --- | --- | --- | --- | --- | --- | --- | --- | --- | --- | --- |
| Pool | Retrieval | Feedback | | | | | | | | |
| ✗ | ✓ | ✓ | 63.4 | 14.8 | 72.7 | 3.7 | 59.8 | 4 | 75.8 | 6.4 |
| ✓ | ✗ | ✓ | 74.6 | 3.6 | 70.2 | 6.2 | 61.9 | 1.9 | 78.9 | 3.3 |
| ✓ | ✓ | ✓ | 78.2 | | 76.4 | | 63.8 | | 82.2 | |
| | | | **Code Llama-13B** | | | | | | | |
| ✗ | ✗ | ✗ | 45.3 | 20.5 | 41.5 | 21.3 | 35.7 | 16.5 | 49.2 | 23.6 |
| ✓ | ✗ | ✗ | 55.8 | 10 | 48.9 | 13.9 | 42.4 | 9.8 | 57.6 | 15.2 |
| ✗ | ✓ | ✗ | 49.6 | 16.2 | 53.2 | 9.6 | 39.5 | 12.7 | 54.1 | 18.7 |
| ✗ | ✗ | ✓ | 47.2 | 18.6 | 50.1 | 12.7 | 41.9 | 10.3 | 58.3 | 14.5 |
| ✓ | ✓ | ✗ | 61.4 | 4.4 | 58.7 | 4.1 | 47.3 | 4.9 | 66.5 | 6.3 |
| ✗ | ✓ | ✓ | 52.7 | 13.1 | 60.4 | 2.4 | 49.6 | 2.6 | 69.1 | 3.7 |
| ✓ | ✗ | ✓ | 63.1 | 2.7 | 59.3 | 3.5 | 50.5 | 1.7 | 71.2 | 1.6 |
| ✓ | ✓ | ✓ | 65.8 | | 62.8 | | 52.2 | | 72.8 | |

There are significant differences in the performance improvements of models when using different combinations of prompting strategies. In the three representative models (GPT-4, PaLM-540B, Code Llama-13B), the complete strategy combination (i.e., with the shared information pool, retrieval, and feedback mechanisms all enabled) typically yields the best results. For example, with GPT-4, the matchability reaches 97.2, executability and accuracy are 92.4 and 86.8, respectively, and readability is 93.4, all of which significantly outperform other combinations. This suggests that integrating multi-layered prompting strategies substantially improves the accuracy and consistency of geospatial code generation.When analyzing the performance of individual strategies, there is also a consistent pattern across different models. For large models, such as GPT-4, enabling the retrieval mechanism alone resulted in a notable improvement in executability from 27.9 to 74.8 and a 20.8 percentage point increase in matchability. This indicates that the retrieval mechanism helps the model effectively access external information in complex tasks, enhancing both code execution and matching performance. In contrast, smaller models, such as Code Llama-13B, show more significant improvements when the feedback mechanism is enabled. For instance, executability increased from 41.5 to 50.1, and matchability improved from 45.3 to 47.2, suggesting that smaller models rely more on the feedback mechanism for external corrections to compensate for their limited knowledge representation.In the pairwise strategy combinations, using GPT-4 as an example, the combination of the shared information pool and the retrieval mechanism achieved a matchability of 89.5 and executability of 85.2, demonstrating that this combination effectively supports the coherence and executability of code when integrating multi-dimensional geospatial data. In contrast, the combination of the retrieval and feedback mechanisms was more advantageous in terms of execution reliability, improving GPT-4's executability to 88.9 and accuracy to 80.3, indicating that retrieval helps the model dynamically obtain external data, while the feedback mechanism corrects errors in real-time during the generation process. The combination of the shared information pool and feedback mechanism performed best in terms of readability, with GPT-4's readability improving to 91.4. This suggests that the shared information pool provides a consistent contextual background, while the feedback mechanism optimizes the output structure and logical clarity. The specific numerical results are visualized in Figure 10.

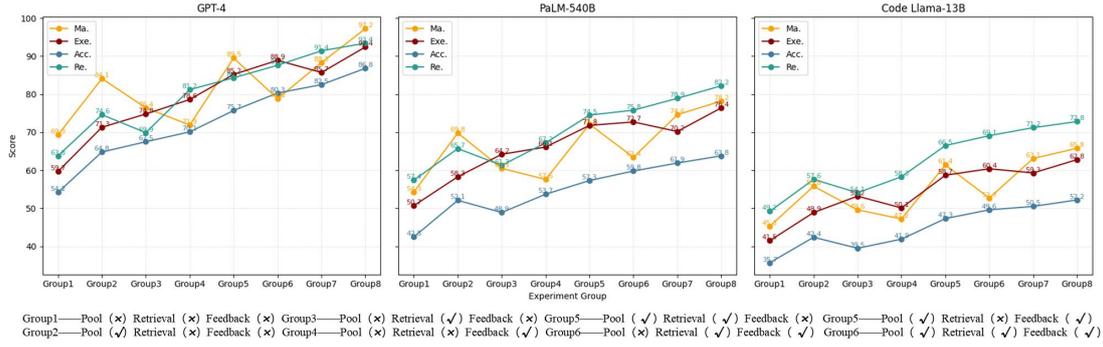

Figure 10 Performance Improvements of Different Strategy Combinations

### 4.3.4. EQ4: Comparison of the Effectiveness of Code Debugging Iterations?

During the code testing and maintenance phase, the generated code is tested. If the code fails to execute correctly or the results do not meet expert expectations, error messages from the console or outputs that do not match expectations can be fed into the CoP process for further correction. In the CoP process, the maximum number of maintenance iterations can be controlled via hyperparameter settings. This study set different maintenance frequencies, including no maintenance, a maximum of 1, 3 (default setting), and 5 iterations, to observe the impact on executability (i.e., whether the code can run properly) and correct execution rate (i.e., whether the code not only runs but also produces expected outputs). It is important to note that "no maintenance" refers to no final maintenance within the CoP process, rather than zero-shot execution. The specific scores for EQ4 are shown in Table 8.

Table8 Scores for Evaluation Metric EQ4

| Strategy | Exe. | | | Acc. | | |
|---|---|---|---|---|---|---|
| | zero-shot | CoP | difference | zero-shot | CoP | difference |
| GPT-4 | | | | | | |
| **Debugging@0** | 52.8 | 66.2 | 13.4 | 42.6 | 50.8 | 8.2 |
| **Debugging@1** | 64.4 | 74.8 | 10.4 | 50.6 | 61.4 | 10.8 |
| **Debugging@3** | **72.9** | **92.4** | **19.5** | **59.2** | **86.8** | **27.6** |
| **Debugging@5** | 73.6 | 94.2 | 20.6 | 61.8 | 87.4 | 25.6 |
| PaLM-540B | | | | | | |
| **Debugging@0** | 47.2 | 61.8 | 14.6 | 40.2 | 47.4 | 7.2 |
| **Debugging@1** | 56.4 | 68.6 | 12.2 | 47 | 55.8 | 8.8 |
| **Debugging@3** | **65.8** | **76.4** | **10.6** | **53.6** | **63.8** | **10.2** |
| **Debugging@5** | 69.2 | 78.4 | 9.2 | 55.4 | 67.6 | 12.2 |
| **Code Llama-13B** | | | | | | |
| **Debugging@0** | 23.2 | 55.2 | 32 | 17.2 | 42.2 | 5.8 |
| **Debugging@1** | 43.4 | 61.2 | 17.8 | 42.8 | 50.4 | 7.6 |
| **Debugging@3** | **58.6** | **62.8** | **4.2** | **46.8** | **52.2** | **5.4** |
| **Debugging@5** | 60.2 | 63.4 | 3.2 | 48.2 | 54.8 | 6.6 |

The results indicate that code maintenance significantly improves both the executability and correct execution rate of the code, validating the necessity of the maintenance step in the CoP process. As the number of maintenance iterations increases, there is a notable improvement in both executability (Exe.) and correct execution rate (Acc.). For example, with GPT-4, executability increased from 66.2% at Debugging@0 to 94.2% at Debugging@5, and the correct execution rate improved from 50.8% to 87.4%. This demonstrates that multiple maintenance iterations effectively enhance the probability of correct code execution. In contrast, the zero-shot strategy performed significantly worse than using the CoP process, further validating the CoP process's role in improving code generation quality. Across all models, regardless of the number of maintenance iterations, executability and correct execution rates under the CoP process were consistently higher than those with the zero-shot strategy. For instance, with GPT-4, the zero-shot executability was 52.8%, but it rapidly increased to 74.8% after Debugging@1. Similarly, the correct execution rate of Code Llama-13B rose from 17.2% under zero-shot to 50.4% after Debugging@1.The maintenance gains vary with model size. Larger models (e.g., GPT-4, PaLM-540B) showed more significant improvements in maintenance. For instance, with GPT-4, from Debugging@3 to Debugging@5, executability increased by 1.8%, and the correct execution rate improved by 1.2%. In contrast, smaller models (e.g., Code Llama-13B) showed smaller gains, with executability increasing by only 0.6% and correct execution rate improving by 2.6% from Debugging@3 to Debugging@5. This suggests that larger models respond more strongly to multiple maintenance iterations, while smaller models experience diminishing returns in maintenance, likely due to the limitations in their parameter size, which restricts their ability to respond to complex corrections.Additionally, the gain differences show that the initial maintenance step (e.g., from Debugging@0 to Debugging@1) typically yields the largest improvement, with subsequent gains gradually decreasing. For example, Code Llama-13B saw a 32 percentage point increase in executability from zero-shot to Debugging@0, while the gain from Debugging@3 to Debugging@5 was only 3.2 percentage points. This demonstrates that the marginal benefit of initial maintenance is the highest, and subsequent maintenance yields diminishing returns, providing a basis for optimizing the maintenance frequency.

**5. Case Study**

To comprehensively demonstrate the effectiveness of the CoP strategy in generating geospatial code, this study selects two representative geospatial code generation scenarios for case analysis. These two scenarios are representative across multiple dimensions: the first case study region is located in China, while the second is in an international region; one case focuses on large-scale, coarse-grained mapping at the provincial level, while the other centers on fine-grained mapping at the level of urban block buildings. In terms of data sources, the former case uses data downloaded from open-source platform interfaces, while the latter relies on locally sourced data. Regarding the execution environment, the former can be run in a local compilation environment, while the latter depends on a cloud-based analysis platform. In terms of data processing, one case involves data reading, indicator sorting, and image annotation, while the other uses independent visualization to present the distribution of analyzed indicator data.Additionally, the two cases each illustrate the processes of the "User Requirement Supplementation" and "Code Executability Confirmation" mechanisms, respectively. These multidimensional demonstrations effectively showcase the applicability and validity of the CoP strategy in geospatial code generation tasks for LLMs.

## 5.1. Local Platform

This case study utilizes the Python platform, combined with libraries such as OSMnx, GeoPandas, and Matplotlib, to retrieve building data for nine designated areas in Manhattan, New York City, from OpenStreetMap. The goal is to generate high-resolution building distribution maps, showcasing the spatial distribution of buildings in each area and highlighting the ten largest buildings by area. The analysis covers nine representative neighborhoods in Manhattan, including the Financial District, Tribeca, SoHo, Chinatown, the Lower East Side, the East Village, Greenwich Village, Chelsea, and the Upper West Side. This analysis uses static building data, with no specific time span involved.In terms of data processing, building data is first retrieved for each area, the area of each building is calculated, and the largest buildings are selected. A unique color scheme is applied to each area to ensure the image boundaries are square-shaped. The final output is a high-resolution JPG image at 300 DPI, visually presenting the building distribution across different areas. This provides valuable support for geospatial analysis and urban planning, as illustrated in Figure 11.

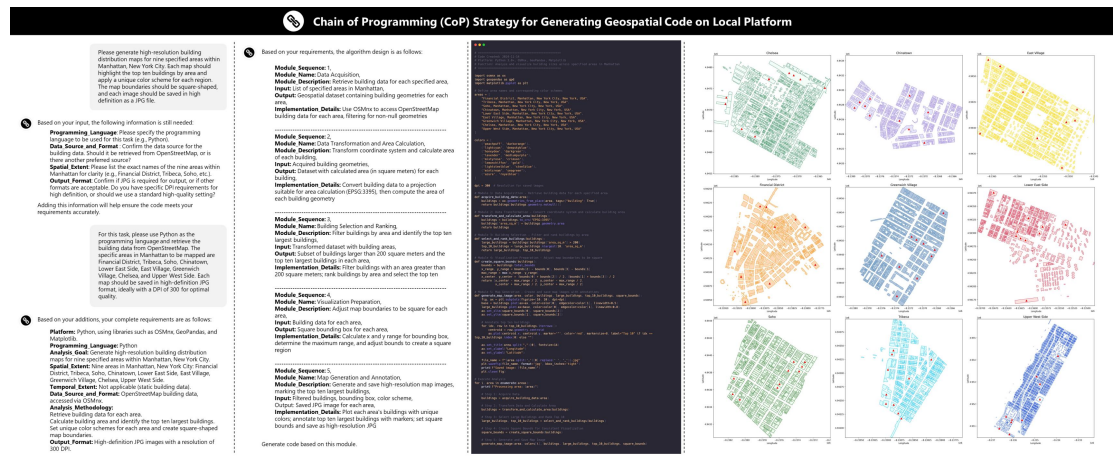

Figure 11 Building Distribution Maps of Nine Manhattan Neighborhoods

## 5.2. Cloud Platform

This case study is based on the Google Earth Engine platform, utilizing JavaScript code to analyze the burned area in Henan Province for the year 2022. The goal is to calculate the total burned area across the province and further break it down by city-level burned areas. To achieve this, the study uses MODIS fire data provided by Google Earth Engine, identifying burned pixels for the specified year and calculating the corresponding burned area (in square kilometers). The analysis covers the entire Henan Province and presents the results at the city-level.The MODIS fire data is filtered within the geographic boundaries of Henan Province for the specified year, and a burned pixel mask is created to identify the burned regions. The burned pixels are then statistically analyzed to calculate the burned area, which is aggregated and presented by city. For visualization, the study displays the burned intensity across different regions on a map, using a color gradient to represent the severity of burning, with labels on the most severely affected areas. Additionally, a pie chart is generated to visually show the proportion of burned area in each city. The final output includes map layers (with color gradients indicating burned intensity), a pie chart illustrating the burned area proportions by city, and a textual summary containing the total number of burned pixels, resolution, and burned area (in square kilometers), as shown in Figure 12.

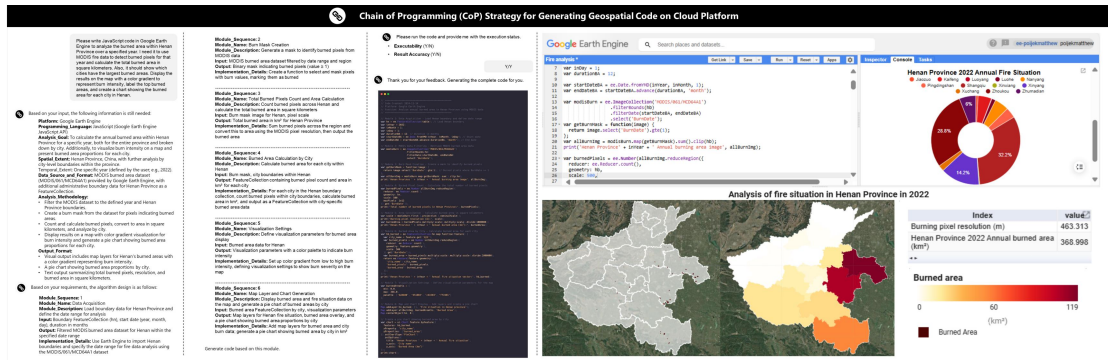

Figure 12 Burned Area Distribution and Proportions in Henan Province (2022)

## 6. Conclusion

This paper introduces the Chain-of-Processing (CoP) programming framework, which builds an end-to-end pipeline from requirement analysis, algorithm design, code implementation, testing, and maintenance, aimed at overcoming the "coding hallucination" problem currently faced by LLMs in geospatial code generation tasks. The framework enhances the logical clarity, syntactical accuracy, and executability of generated code through innovative mechanisms such as a shared information pool, knowledge base retrieval, and expert feedback. Through case studies on building data visualization and fire damage analysis, the applicability and effectiveness of the CoP framework in various geospatial applications are validated, providing a more systematic and practical solution for geospatial modeling.

Despite the significant advantages demonstrated by the CoP framework in geospatial code generation, the shared information pool and knowledge base retrieval mechanisms still require further optimization to enhance compatibility with different data sources and platforms. Additionally, the current expert feedback mechanism primarily relies on human intervention, and future research could explore automated feedback models to further improve code generation efficiency. As the complexity of geospatial data and application demands continue to grow, the expansion of CoP strategies into other vertical domains warrants further investigation, particularly in automating the transition from models to code, offering new insights for cross-domain code generation with LLMs.

**Disclosure statement**

No potential conflict of interest was reported by the authors.

**Funding**

The work was supported by National Natural Science Foundation of China (no. 41930107).

**Data availability statement**

Data are available on request from the authors. The data that support the findings of this study are available from thecorresponding author, Huayi Wu, upon reasonable request.